\documentclass[prl,amssymb,nobibnotes,superscriptaddress,longbibliography,notitlepage,twocolumn]{revtex4-1}

\usepackage{enumerate,amsmath}
\usepackage[dvipdfmx]{graphicx}

\begin{document}

\title{Power of one non-clean qubit}%
\author{Tomoyuki Morimae}
\email{morimae@gunma-u.ac.jp}
\affiliation{ASRLD Unit, Gunma University, 1-5-1 Tenjincho, Kiryu,
Gunma, 376-0052, Japan}
\author{Keisuke Fujii}
\email{fujii@qi.t.u-tokyo.ac.jp}
\affiliation{Photon Science Center, Graduate School of Engineering, 
The University of Tokyo, 2-11-16 Yayoi, Bunkyoku, Tokyo 113-8656, Japan}
\affiliation{JST, PRESTO, 4-1-8 Honcho, Kawaguchi, Saitama,
332-0012, Japan}
\author{Harumichi Nishimura}
\email{hnishimura@is.nagoya-u.ac.jp}
\affiliation{Graduate School of Information Science,
Nagoya University, Furocho, Chikusaku, Nagoya, Aichi, 464-8601, Japan}

\date{\today}
\begin{abstract}
The one-clean qubit model (or the DQC1 model)
is a restricted model of quantum computing
where only a single qubit of the initial state is pure
and others are maximally mixed.
Although the model is not universal, 
it can efficiently solve several problems whose classical
efficient solutions are not known. Furthermore, it was recently
shown that if the one-clean qubit model is classically efficiently
simulated, the polynomial hierarchy collapses to the
second level.
A disadvantage of the one-clean qubit model is, however,
that the clean qubit is too clean: for example,
in realistic NMR experiments,
polarizations are not enough high to have the perfectly pure qubit.
In this paper, we consider a more realistic one-clean qubit model,
where the clean qubit is not clean, but depolarized.
We first show that, for any polarization,
a multiplicative-error calculation of
the output probability distribution of the model is possible in
a classical polynomial time if we take an appropriately large
multiplicative error. The result is in a strong contrast to
that of the ideal one-clean qubit model
where the classical efficient multiplicative-error
calculation (or even the sampling) with the same
amount of error causes the
collapse of the polynomial hierarchy.
We next show that, for any polarization lower-bounded
by an inverse polynomial, 
a classical efficient sampling 
(in terms of a sufficiently small
multiplicative error or an exponentially-small
additive error) of the output
probability distribution of the model is impossible
unless BQP is contained
in the second level of the polynomial hierarchy,
which suggests the hardness of the classical efficient simulation
of the one non-clean qubit model.
\end{abstract}

\maketitle

To show a supremacy of quantum computing over classical one
is one of the most central research subjects in physics and computer science.
Although several quantum advantages have been shown in terms of
the communication complexity~\cite{comm_complex1,comm_complex2} 
and the query complexity~\cite{Grover,Simon}, the ultimate question
``is ${\rm BPP}\neq{\rm BQP}$?"
remains open.

One good strategy to study the gap between quantum and classical
is restricting the quantum side. It is also important from
the experimental point of view given the high technological
demands for the realization of a universal quantum computer.
For example, quantum computing that uses
only Clifford gates~\cite{Gottesman,GottesmanAaronson} 
or Fermionic linear optical 
gates (or the matchgates)~\cite{Valiant,Terhal,Knill,Jozsa} 
is classically efficiently simulatable.
On the other hand, restricted models that do not seem to be
classically efficiently simulatable do exist.
For example, 
if quantum computing that uses only non-interacting 
Bosons~\cite{BS} or commuting gates~\cite{IQP,IQP2,IQP3} 
(so called the IQP model) is classically
efficiently simulated, then the polynomial hierarchy collapses
to the third level (or the second level~\cite{FKMNTT}). 
Since a collapse of the polynomial hierarchy is not believed to
happen, these results suggest the hardness of 
the classical efficient simulation of these restricted models.

The one-clean qubit model (or the DQC1 model)~\cite{KnillLaflamme}
is another restricted model of quantum computing that
is believed to be stronger than classical computing.
The model was originally motivated by NMR, 
which has over half a century history and
matured control schemes~\cite{ChuangNMR,NMR1,NMR2}.
An NMR spin ensemble system has several 
physical advantages:
for example, molecules consisting of wide varieties of nuclear and 
electron spins can be chemically synthesized.
Furthermore, the macroscopic signals are obtained by virtue of the 
huge number of copies in the ensemble with less backaction. 
Finally, each spin is highly isolated from 
external degrees of freedom, which is favorable 
to avoid decoherence. 
Because of these reasons,
an NMR spin ensemble system is a useful experimental setup 
to probe quantum many-body dynamics,
and in fact, it has been applied to several quantum information processing
tasks including the quantum simulation~\cite{NMRexp}.
However, the low decoherence rate is a double-edged sword:
in NMR, an initialization or polarization of a nuclear spin
is not easy.
Therefore, NMR quantum information processing 
has to be a highly mixed state quantum computation.

The one-clean qubit model formalizes the NMR quantum information
processing in the following way:
First, the initial state is 
$
|0\rangle\langle0|\otimes(\frac{I}{2})^{\otimes n-1},
$
where $I\equiv|0\rangle\langle0|+|1\rangle\langle1|$
is the two-dimensional identity operator.
Second, any (uniformly-generated
polynomial-time) $n$-qubit unitary operator is applied on it.
Finally, some qubits are measured in the computational basis.
(Note that in some strict definitions, only a single-qubit is
allowed to be measured, or only an expectation value
of a single-qubit measurement is obtained.) 

If the clean qubit $|0\rangle$ of the initial state
is replaced with the maximally-mixed state $\frac{I}{2}$, 
the quantum computing
is trivially simulatable with a polynomial-time classical computer,
since 
$
U(\frac{I}{2})^{\otimes n}U^\dagger=
(\frac{I}{2})^{\otimes n}
$
for any unitary operator $U$.
This example suggests that 
the one-clean qubit model is also classically
efficiently simulatable,
since only a single pure qubit does not seem to cause
any drastic change.
However, surprisingly, the model 
can
efficiently solve several problems
whose classical efficient solutions are not known, 
such as the spectral density estimation~\cite{KnillLaflamme}, 
testing integrability~\cite{integrability}, calculations of 
the fidelity decay~\cite{fidelity_decay}, and approximations of the Jones 
and HOMFLY polynomials~\cite{ShorJordan08,Passante09,JordanWocjan09}.
Furthermore, it was recently shown that
if the probability distribution of the measurement result
on the single output qubit of the one-clean qubit model
is classically
efficiently sampled (in terms of a multiplicative error
or an exponentially-small additive error),
then
the polynomial hierarchy collapses to
the second level~\cite{MorimaeFF,FKMNTT}.

A disadvantage of the one-clean qubit model is, however,
that the clean qubit
is too clean:
for example,
in realistic experiments,
the polarization of spins in an NMR ensemble 
is not high enough to obtain the perfectly pure qubit
(even if the algorithmic cooling or the quantum data compression
~\cite{DataCompress,Boykin,CleveDiVincenzo} is employed).
Therefore, the following important question remains open:
can we show any hardness of a classical efficient simulation
of a more realistic one ``non-clean" qubit model?

In this paper, we consider a modified version
of the one-clean qubit model
where the clean qubit of the initial state
is not clean but depolarized
(Eq.~(\ref{initial})).
We first show that for any polarization,
a multiplicative-error calculation
of the output probability distribution of the model is possible
in a classical polynomial time if we take a sufficiently large
multiplicative error.
Note that the result is in a strong contrast to that
of the ideal one-clean qubit model where
the classical efficient
multiplicative-error calculation (or even the sampling)
with the same amount of error
causes the collapse of the polynomial hierarchy~\cite{MorimaeFF,FKMNTT}.
We also point out that the bound of the multiplicative error
is optimal by showing a counter example
for errors smaller than the bound.
We next consider the sampling of the output probability
distribution of our model. We show that
for any polarization lower-bounded by an inverse polynomial,
a classical efficient sampling (in terms of
a sufficiently small
multiplicative error or an exponentially-small additive error)
is impossible unless
BQP is contained in the second level of the polynomial hierarchy.
Since it is not believed to happen~\cite{AaronsonPH},
the result demonstrates the power of
one non-clean qubit.

Note that, with similar and other motivations,
noisy versions of IQP circuits
have been studied recently, and shown to
be hard to classically efficiently simulate~\cite{IQP3,FujiiTamate}.
Moreover, quantum computing that uses a universal gate set
but is too noisy to realize fault-tolerant universal quantum computing
was shown to be hard to classically efficiently simulate~\cite{Fujii_recent}.

{\it One non-clean qubit model}.---
We consider the following model.
The initial state is the $n$-qubit state
\begin{eqnarray}
\rho_\epsilon^{init}&\equiv&\left(\frac{1+\epsilon}{2} |0\rangle \langle 0|
+ \frac{1-\epsilon}{2} |1\rangle \langle 1| \right)
\otimes \left(\frac{I}{2} \right)^{\otimes (n-1)},
\label{initial}
\end{eqnarray}
where 
the first qubit corresponds to 
the nuclear spin to be probed
whose polarization $\epsilon$ is relatively higher 
than the others but still very small.
The case $\epsilon=1$ corresponds to the original
one-clean qubit model.
Any (uniformly-generated polynomial-time) 
$n$-qubit unitary operator $U$ is applied
on the initial state to obtain
$\rho_\epsilon\equiv U\rho_\epsilon^{init}U^\dagger$.
Finally some qubits are
measured in the computational basis.
If we measure all qubits,
the probability $p_z$ of obtaining the result $z\in\{0,1\}^n$
is
\begin{eqnarray*}
p_z\equiv\langle z|\rho_\epsilon|z\rangle
=\epsilon\langle z|
U\Big(|0\rangle\langle0|\otimes\frac{I^{\otimes (n-1)}}{2^{n-1}}\Big)
U^\dagger|z\rangle+\frac{1-\epsilon}{2^n}.
\end{eqnarray*}


{\it Multiplicative-error calculation}.---
First, we consider calculations of the output probability distribution
of the model. As is shown in Appendix A,
the exact calculation is trivially $\#$P-hard
(actually GapP-complete). We therefore consider approximations,
namely, multiplicative-error calculations.
Here, a multiplicative-error approximation with
the error $c\ge0$ means that
the target value $p$ and the calculated value $q$ satisfy
$|p-q|\le cp$.

{\bf Result 1}:
For any $0\le\epsilon<1$,
$p_z$ can be approximated by the uniform
distribution $q_z=\frac{1}{2^n}$ with any multiplicative error
$c$ that satisfies $c\ge\frac{\epsilon}{1-\epsilon}$.

{\bf Proof}:
We can show 
\begin{eqnarray*}
\frac{1-\epsilon}{2^n}\le p_z\le \frac{1+\epsilon}{2^n}
\end{eqnarray*}
for any $z\in\{0,1\}^n$. Therefore,
\begin{eqnarray*}
\Big|p_z-\frac{1}{2^n}\Big|\le\frac{\epsilon}{2^n}
=\frac{\epsilon}{1-\epsilon}\frac{1-\epsilon}{2^n}
\le cp_z.
\end{eqnarray*}
\fbox

According to the result of Ref.~\cite{FKMNTT}, 
if the output probability distribution of
the computational-basis measurement on the single
output qubit of the one-clean qubit model
is classically efficiently sampled with
the $c=1-\frac{1}{2^n}$ multiplicative-error,
then the polynomial hierarchy collapses to the second 
level. (See Appendix B.)
Result 1 shows that the hardness result does no longer hold
for the one non-clean qubit case.
In fact, from Result 1, for any $x\in\{0,1\}$,
\begin{eqnarray*}
&&\Big|\sum_{y\in\{0,1\}^{n-1}}p_{xy}-\frac{1}{2}\Big|
=
\Big|\sum_{y\in\{0,1\}^{n-1}}p_{xy}-\sum_{y\in\{0,1\}^{n-1}}\frac{1}{2^n}\Big|\\
&\le&
\sum_{y\in\{0,1\}^{n-1}}\Big|p_{xy}-\frac{1}{2^n}\Big|
\le
c\sum_{y\in\{0,1\}^{n-1}}p_{xy},
\end{eqnarray*}
which means that the probability
$\sum_{y\in\{0,1\}^{n-1}}p_{xy}$ of obtaining $x\in\{0,1\}$
when the first qubit of our model is measured in the computational
basis is approximated to $\frac{1}{2}$ (and therefore
classically efficiently sampled) with the multiplicative
error $c$.
If we take the polarization $\epsilon$ as
$\epsilon\le\frac{1}{2}-\frac{1}{2^{n+2}-2}$,
for example, $c$ can be $c=1-\frac{1}{2^n}$.

{\it Optimality of the bound}.---
We can show that the bound
$c\ge\frac{\epsilon}{1-\epsilon}$ of Result 1
is optimal in the following sense:

{\bf Result 2}:
For any $0\le \epsilon<1$ and $c\ge0$ such that
$0\le c<\frac{\epsilon}{1-\epsilon}$,
and for any probability distribution 
$q:\{0,1\}^n\ni z\mapsto q_z\in[0,1]$,
there exists an $n$-qubit unitary operator $U$ such that
$
|p_z-q_z|>cp_z
$
for a certain $z\in\{0,1\}^n$.

{\bf Proof}:
If $q_z<\frac{1}{2^n}$ for all $z\in\{0,1\}^n$, 
then $\sum_{z\in\{0,1\}^n}q_z<1$, which is contradiction.
Therefore there is at least one 
$y\in\{0,1\}^n$ such that $q_y\ge\frac{1}{2^n}$.
Let $y_1\in\{0,1\}$ be the first bit of $y$.
If we take 
$U=X^{y_1\oplus1}\otimes I^{\otimes n-1}$,
\begin{eqnarray*}
p_y&=&\epsilon\langle y|
U\Big(|0\rangle\langle0|\otimes\frac{I^{\otimes n-1}}{2^{n-1}}\Big)
U^\dagger|y\rangle
+\frac{1-\epsilon}{2^n}\\
&=&\epsilon\langle y|
\Big(|y_1\oplus1
\rangle\langle y_1\oplus1|\otimes\frac{I^{\otimes n-1}}{2^{n-1}}\Big)
|y\rangle
+\frac{1-\epsilon}{2^n}
=\frac{1-\epsilon}{2^n},
\end{eqnarray*}
and therefore
$|p_y-q_y|\ge \frac{\epsilon}{2^n}$,
while
$
cp_y<\frac{\epsilon}{1-\epsilon}\frac{1-\epsilon}{2^n}
=\frac{\epsilon}{2^n}.
$
Hence we obtain $|p_y-q_y|>cp_y$.
\fbox

{\it Multiplicative-error sampling}.---
We next consider the sampling. We first show
the hardness result for the multiplicative error case.

{\bf Result 3}:
Let us assume that
for any (uniformly-generated polynomial-time) 
$n$-qubit unitary operator $U$,
there exists
a $poly(n)$-time classical probabilistic algorithm
that outputs $z\in\{0,1\}^n$ with probability $q_z$
such that 
\begin{eqnarray}
|p_{0^n}-q_{0^n}|\le cp_{0^n}
\label{ass_multi}
\end{eqnarray}
with a certain $c$ that satisfies
$
0\le c\le\epsilon-\frac{1}{\delta(n)}
$
for a polynomial $\delta>0$.
Then 
BQP is contained in SBP.

Before giving a proof, there are five remarks.
First, the value of $c$ considered in Result 3 
is always smaller than that of Result 1,
since $\frac{\epsilon}{1-\epsilon}-(\epsilon-\frac{1}{\delta})\ge0$,
and therefore there is no contradiction between these two results.
Second, since $\frac{\epsilon}{1-\epsilon}=\epsilon+O(\epsilon^2)$
for small $\epsilon$, the combination of Result 1
and Result 3 roughly means that $\epsilon$ is
the threshold for $c$: if $c>\epsilon$ then the classical simulation
is possible (Result 1), while if $c<\epsilon$ then it is 
impossible (Result 3). 
Third, 
Result 3 implicitly assumes that $\epsilon$ is lower-bounded
by an inverse polynomial, since otherwise no
$c$ can satisfy $c\le \epsilon-\frac{1}{\delta}$.
The assumption, $\epsilon\ge1/poly$, is acceptable, since
we can take such $\epsilon$ in realistic NMR experiments.
(Actually, $\epsilon$ can be even a small
but system-size-independent constant.)
Fourth, the standard definition of the multiplicative-error 
sampling is that
$|p_z-q_z|\le cp_z$ for any $z\in\{0,1\}^n$, but in Result 3,
the satisfiability only for $z=0^n$ is enough. 
Finally, SBP is defined in
the following way~\cite{SBP}:
A language $L$ is in SBP if there exist a polynomial $r$
and a uniformly-generated
family of polynomial-size probabilistic
classical circuits
such that
if $x\in L$ then the acceptance probability is $\ge 2^{-r(|x|)}$,
and
if $x\notin L$ then the acceptance probability is $\le 2^{-r(|x|)-1}$.
As is shown in Appendix C, 
the bound $(2^{-r},2^{-r-1})$
can be replaced with 
$(a2^{-r},b2^{-r})$ for any $0\le b<a\le 1$ such
that $a-b\ge \frac{1}{poly}$.
It is known that SBP is in AM~\cite{SBP},
and therefore in the second level of the polynomial hierarchy:
$
{\rm SBP}\subseteq{\rm AM} \subseteq\Pi_2^p.
$
Hence ${\rm BQP}\subseteq{\rm SBP}$ means that BQP is in the second level
of the polynomial hierarchy.
Note that ${\rm BQP}\subseteq{\rm SBP}$ itself is also unlikely,
since ${\rm SBP}\subseteq{\rm BPP}_{\rm path}$
and there is an oracle such that
${\rm BQP}$ is not contained in ${\rm BPP}_{\rm path}$~\cite{Chen}.

{\bf Proof}:
Let us assume that a language $L$ is in BQP.
This means that for any polynomial $r$,
there exists a uniformly generated
family $\{V_x\}$ of polynomial-size quantum circuits such that
\begin{eqnarray*}
\langle 0^n|V_x^\dagger
(|0\rangle\langle0|\otimes I^{\otimes n-1})V_x|0^n\rangle
\left\{
\begin{array}{ll}
\ge 1-2^{-r}&(x\in L)\\
\le 2^{-r}&(x\notin L).
\end{array}
\right.
\end{eqnarray*}
Here, $n=poly(|x|)$.
Let us take $U=V_x^\dagger$.
We also take $r$ such that $\epsilon2^{-r+1}\le\frac{1}{2\delta}$.

If $x\in L$,
\begin{eqnarray*}
q_{0^n}&\ge&
(1-c)\Big[
\frac{\epsilon}{2^{n-1}}
\langle0^n|
V_x^\dagger
(|0\rangle\langle0|\otimes I^{\otimes n-1})V_x
|0^n\rangle+\frac{1-\epsilon}{2^n}
\Big]\\
&\ge&
(1-c)\Big[
\frac{\epsilon}{2^{n-1}}
(1-2^{-r})
+\frac{1-\epsilon}{2^n}
\Big]\\
&=&
\frac{(1-c)}{2^n}(
1+\epsilon-\epsilon2^{-r+1}
).
\end{eqnarray*}
If $x\notin L$,
\begin{eqnarray*}
q_{0^n}&\le&
(1+c)\Big[
\frac{\epsilon}{2^{n-1}}
\langle0^n|
V_x^\dagger
(|0\rangle\langle0|\otimes I^{\otimes n-1})V_x
|0^n\rangle+\frac{1-\epsilon}{2^n}
\Big]\\
&\le&
(1+c)\Big[
\frac{\epsilon}{2^{n-1}}
2^{-r}
+\frac{1-\epsilon}{2^n}
\Big]\\
&=&
\frac{(1+c)}{2^n}(
1-\epsilon
+\epsilon2^{-r+1}
).
\end{eqnarray*}
Since
\begin{eqnarray*}
&&(1-c)(1+\epsilon-\epsilon2^{-r+1})
-
(1+c)(1-\epsilon+\epsilon2^{-r+1})\\
&=&
2(\epsilon-\epsilon2^{-r+1}-c)\\
&\ge&2\Big(\epsilon-\frac{1}{2\delta}-
\Big(\epsilon-\frac{1}{\delta}\Big)\Big)=\frac{1}{\delta},
\end{eqnarray*}
$L$ is in SBP.
\fbox

{\it Exponentially-small additive error sampling}.---
We can also show a similar hardness result
for the exponentially-small additive error case.

{\bf Result 4}:
Let us assume that for any
(uniformly-generated polynomial-time) $n$-qubit unitary operator $U$,
there exists 
a $poly(n)$-time classical probabilistic algorithm
that outputs $z\in\{0,1\}^n$ with probability $q_z$
such that 
\begin{eqnarray}
|p_{0^n}-q_{0^n}|\le\eta
\label{ass_add}
\end{eqnarray}
with a certain $\eta$ that satisfies 
$0\le\eta\le(\epsilon-\frac{1}{\delta})2^{-n}$
for a polynomial $\delta>0$.
Then BQP is contained in SBP.

Before giving a proof, there are two remarks:
First, we again implicitly assume $\epsilon\ge1/poly$.
Second, the assumption Eq.~(\ref{ass_add}) can be replaced with
the more standard assumption ($L_1$-norm additive-error approximation),
$
\sum_{z\in\{0,1\}^n}|p_z-q_z|\le\eta,
$
since if it is satisfied then
$
|p_{0^n}-q_{0^n}|\le
\sum_{z\in\{0,1\}^n}|p_z-q_z|\le\eta,
$
and therefore Eq.~(\ref{ass_add}) is satisfied.
Since Eq.~(\ref{ass_add}) is weaker, we have used it.

{\bf Proof}:
Let us assume that a language $L$ is in BQP,
and let $V_x$ be the corresponding circuit
as assumed in the proof of Result 3.
We take $U=V_x^\dagger$, and $r$ such that
$\epsilon2^{-r+1}\le\frac{1}{2\delta}$.
If $x\in L$,
$
q_{0^n}\ge
\frac{1}{2^n}(1+\epsilon-\epsilon2^{-r+1}-
2^n\eta).
$
If $x\notin L$,
$
q_{0^n}\le
\frac{1}{2^n}(2^{-r+1}\epsilon+1-\epsilon
+2^n\eta
).
$
Since
\begin{eqnarray*}
&&(1+\epsilon-\epsilon2^{-r+1}-2^n\eta)
-(1-\epsilon+\epsilon2^{-r+1}+2^n\eta)\\
&\ge&2\Big(\epsilon-\frac{1}{2\delta}-\Big(\epsilon-\frac{1}{\delta}\Big)
\Big)=\frac{1}{\delta},
\end{eqnarray*}
$L$ is in SBP.
\fbox

{\it Discussion}.---
In this paper, we have used a multiplicative or
an exponentially-small additive error in the definition
of the classical samplability.
It is an important open problem whether we can generalize 
the results to a
constant or inverse-polynomial $L_1$-norm error
as was done for the Boson sampling~\cite{BS},
the IQP~\cite{IQP2,IQP3}, and the Fourier sampling~\cite{Fourier}.
(These results do not seem to be directly applied to the 
one qubit model, even in the perfect polarization case,
since the one qubit model
seems to be able to simulate standard quantum computing
with only an exponentially small rate.)
In the present case, however, using a multiplicative or an
exponentially-small
additive error is justified, since 
in our case the model itself is noisy. In other words,
we consider the following sampling problem:
``sample the output probability distribution of a noisy
one-clean qubit model". The problem can be, of course,
exactly solvable with the noisy one-clean qubit model,
but we have shown that solving 
the problem classically
is impossible even with a multiplicative or an exponentially-small
additive error.
We have therefore shown the existence of a sampling problem
that can be exactly solvable by a realistic non-universal
quantum computer but cannot be solved by a classical computer
even with a multiplicative
or an exponentially-small additive error.

We have considered the output probability distribution
of the measurements on all qubits.
It is an open problem whether we can 
reduce the number of measured qubits
to one.
Furthermore,
we want to improve
our consequence,
${\rm BQP}\subseteq{\rm SBP}$, 
to more unlikely one such as the collapse of the polynomial
hierarchy, but at this moment we do not know how to do it.

Finally, to conclude this paper, let us discuss 
roles of entanglement in NMR quantum computing.
In Ref.~\cite{Braunstein99},
a criteria on the initial polarization, below which 
the system becomes a separable state,
was derived, 
and pointed out that states used in
NMR experiments are separable states.
It sounds like NMR quantum information processing
has no quantum power,
and in fact some researchers have insisted that NMR quantum information
processing is useless.
The conclusion is, however, wrong.
For example, as is shown in the present paper,
NMR quantum computing can demonstrate the quantum supremacy
for some sampling problems.
Furthermore, in the first place, entanglement 
is not directly connected to the quantum speedup:
recently it was shown that
a larger entanglement does not
necessarily mean a quantum speedup~\cite{Bremner09,Gross09},
and that
quantum computing whose register is always
bi-separable can solve any BQP problem~\cite{Nest}.
Interestingly, 
even if we consider a much weaker model, which we
call separable quantum computing, where
the register is always separable during the computation,
its classical simulatability is not so obvious.
For example, even if the register is always separable,
it seems to be hard to find a separable decomposition 
after every local unitary gate operation,
since after a local unitary gate operation, some pure states
in the mixture can be entangled.
Furthermore, although any discord free quantum computation 
is classically simulatable~\cite{Eastin},
a separable state can have non-zero quantum discord.

\acknowledgements
TM is supported by Grant-in-Aid for Scientific Research on Innovative Areas
No.15H00850 of MEXT Japan, and the Grant-in-Aid for Young Scientists (B) 
No.26730003 of JSPS. KF is supported by KAKENHI No.16H02211,
PRESTO, JST, CREST, JST, and ERATO, JST. HN is supported 
by the Grant-in-Aid for Scientific Research (A) Nos.26247016 and 16H01705 of 
JSPS, the Grant-in-Aid for Scientific Research on Innovative Areas 
No.24106009 of MEXT, and the Grant-in-Aid for Scientific Research 
(C) No.16K00015 of JSPS.

{\bf Appendix A}.---
It is easy to see that
the exact calculation of the output probability
distribution of our model
is \#P-hard (actually, GapP-complete), because
the ability of the exact calculation of the output probability
distribution of the model allows us to exactly calculate
the output probability distribution of the one-clean qubit model,
which contains (in an exponentially small rate) 
the output probability distribution of any (polynomial-time)
quantum computing. It is known that the exact calculation
of the output probability distribution of (polynomial-time) quantum
computing is \#P-hard (actually GapP-complete)~\cite{FR}.

{\bf Appendix B}.---
Here we show that if the output probability distribution
of the one-clean qubit model is classically efficiently sampled
with the multiplicative error $c=1-\frac{1}{2^n}$ then
NQP is in NP, which causes the collapse of the
polynomial hierarchy to the second level.
We follow the argument in Refs.~\cite{FKMNTT,FKMNTT2}.
Let us assume that a language $L$ is in NQP, 
which means that
there exists 
a uniformly-generated family $\{V_x\}$ of polynomial-size
quantum circuits such that
if $x\in L$ then $0<p<1$, and
if $x\notin L$ then $p=0$, 
where $p$ is the acceptance probability.
It was shown in Ref.~\cite{FKMNTT2} that from $V_x$,
which acts on $n-1$ qubits,
we can construct an $n$-qubit one-clean qubit circuit such that
the probability $\tilde{p}$ of obtaining 1 when the
clean qubit is measured in the computational basis
is $\tilde{p}=\frac{4}{2^{n-1}}p(1-p)$.
Therefore 
if $x\in L$ then $\tilde{p}>0$, and
if $x\notin L$ then $\tilde{p}=0$. 
Let us assume that there exists a classical polynomial-time
probabilistic algorithm whose acceptance probability $q$ satisfies
$|\tilde{p}-q|\le (1-\frac{1}{2^n})\tilde{p}$.
Then,
if $x\in L$ we have $q\ge\frac{\tilde{p}}{2^n}>0$, and
if $x\notin L$ then $q\le(2-\frac{1}{2^n})\tilde{p}=0$. 
Therefore, NQP is in NP, which causes the collapse
of the polynomial hierarchy to the second level.

{\bf Appendix C}.---
Here we show that the bound $(2^{-r},2^{-r-1})$ of SBP
can be replaced with $(a2^{-r},b2^{-r})$
for any $0\le b<a\le 1$ such that $a-b\ge\frac{1}{q}$,
where $q>0$ is a polynomial.

Since $a\ge b+\frac{1}{q}\ge \frac{1}{q}$,
there exists a polynomial $k\ge0$ such that
$a>\frac{1}{2^k}$.
Let $V_x$ be the original circuit of SBP.
We define the modified circuit $V_x'$ in the following way:
it first runs the original circuit $V_x$, and
then accepts with probability $\frac{1}{a2^k}$
if $V_x$ accepts.
If $x\in L$, the acceptance probability of $V_x'$ is
$p_{acc}\ge \frac{a2^{-r}}{a2^k}=\frac{1}{2^{r+k}}$.
If $x\notin L$, it is
\begin{eqnarray*}
p_{acc}&\le& \frac{b2^{-r}}{a2^k}\\
&=&\frac{1}{2^{r+k}}\frac{a-(a-b)}{a}\\
&=&\frac{1}{2^{r+k}}\Big(1-\frac{a-b}{a}\Big)\\
&\le&\frac{1}{2^{r+k}}\Big(1-\frac{1}{q}\Big).
\end{eqnarray*}
We run $V_x'$ $q$ times, and accept if all results
accept.
If $x\in L$,
the acceptance probability is
$
p_{acc}^q\ge
\frac{1}{2^{(r+k)q}}.
$
If $x\notin L$, it is
\begin{eqnarray*}
p_{acc}^q
&\le&\frac{1}{2^{(r+k)q}}\Big(1-\frac{1}{q}\Big)^{q}\\
&=&\frac{1}{2^{(r+k)q}}\Big[\Big(1+\frac{1}{q-1}\Big)^{q}\Big]^{-1}\\
&\le&\frac{1}{2^{(r+k)q}}\frac{1}{2},
\end{eqnarray*}
where we have used
\begin{eqnarray*}
\Big(1+\frac{1}{q-1}\Big)^{q}
&=&\sum_{j=0}^{q}{q\choose{j}}\Big(\frac{1}{q-1}\Big)^j\\
&\ge&1+{{q}\choose{1}}\frac{1}{q-1}\\
&=&1+\frac{q}{q-1}\\
&\ge&2.
\end{eqnarray*}

\end{document}